\newif\ifpreprint       \preprintfalse     
\newif\iflinenumbers    \linenumbersfalse  
\newif\ifcolorlinks     \colorlinkstrue    
\newif\iflongbib        \longbibfalse      
\newif\ifshowkeys       \showkeysfalse     

\ifpreprint
  \documentclass[%
    aps, prb, preprint,
    superscriptaddress,
    nofootinbib,
    floatfix
  ]{revtex4-2}
\else
  \documentclass[%
    aps, prb, reprint,              
    superscriptaddress,
    nofootinbib,
    floatfix
  ]{revtex4-2}
\fi

\usepackage[T1]{fontenc}
\usepackage[utf8]{inputenc}
\usepackage{lmodern}                 
\usepackage{microtype}               
\usepackage{graphicx}                
\usepackage{dcolumn}                 
\usepackage{bm}                      
\usepackage{amsmath,amssymb,mathtools}
\usepackage{siunitx}                 
\usepackage{xspace}                  
\usepackage{physics}                 
\usepackage{orcidlink}               

\usepackage{hyperref}
\ifcolorlinks
  \hypersetup{
    colorlinks=true,
    linkcolor=blue,
    citecolor=blue,
    urlcolor=blue,
    pdfauthor={Your Name},
    pdftitle={Your Title}
  }
\else
  \hypersetup{hidelinks}
\fi

\iflinenumbers
  \usepackage[switch]{lineno}
  \linenumbers
\fi

\ifshowkeys
  \usepackage{showkeys}
\fi

\begin{document}

\title{Quantum Printing: Laguerre-Gaussian Beam Induced Topological Magnetic Textures }


\author{Yuefei Liu}\email{yuefei.liu@su.se}
\affiliation{Nordita, Stockholm
University, and KTH Royal Institute of Technology, Hannes Alfv\'ens vag 12, SE-106 91 Stockholm, Sweden}
\author{Tien-Tien Yeh}
\affiliation{Department of Physics, IMS, University of Connecticut, Storrs, Connecticut 06269, USA}
\author{Alexander V. Balatsky}
\affiliation{Nordita, Stockholm
University, and KTH Royal Institute of Technology, Hannes Alfv´ens vag 12, SE-106 91 Stockholm, Sweden}
\affiliation{Department of Physics, IMS, University of Connecticut, Storrs, Connecticut 06269, USA}

\date{\today}

\begin{abstract}
Structured light has become a practical tool for controlling matter by applying tailored, space- and time-dependent electromagnetic fields.
We show that Laguerre-Gaussian pulses imprint non-collinear magnetic textures via the spatial structure of optical magnetic field. Our route offers a direct spatial selectivity determined by the optical features without relying on material anisotropic interactions. The proposed printing approach does not require interfacial anisotropy or bulk chirality, current-driven torques, or thermal quenching.
We use micromagnetic simulations to demonstrate the potential to create topological charge density emerging during the pulse and reveal control through the optical topological properties and polarization. 
These results suggest structured-light quantum printing as a viable approach for magnonics and motivate studies toward reconfigurable topological textures enabled by ultrafast THz optics and non-thermal control.

\end{abstract}

\maketitle


\section{Introduction}
Topological and noncollinear textures are a unifying theme across quantum many-body systems \cite{nagaosa2013topological,fert2017magnetic}. Quantum printing provides a general nonequilibrium route to engineer such textures by transferring selected photon quantum numbers into collective degrees of freedom of driven quantum matter \cite{qp2025}.
In each case, topology encodes robust, nonperturbative structure in an underlying order parameter, enabling protected responses and collective dynamics that are attractive for both fundamental studies and applications.
A central challenge is to create and control such textures in a fast and spatially programmable manner, ideally without relying on material-specific interactions or large dissipative currents.

Existing routes to generate topological magnetic textures typically rely on intrinsic chiral interactions such as Dzyaloshinskii–Moriya exchange, interfacial anisotropy, or bulk crystal symmetry, which restrict the range of compatible materials~\cite{Bog1994, Ros2006, yu2018transformation, fujita2017encodingoam, gao2024prl}. 
Recent Landau-Lifshitz-Gilbert (LLG) simulations in frustrated magnets show that a linearly polarized AC electric field can stabilize bimeron crystals~\cite{shirato2025}, while Laguerre-Gaussian (LG) beam irradiation can generate skyrmions~\cite{laxamana2026}.
Alternative approaches based on current-induced spin–orbit torques or thermal quenching enable dynamical creation of skyrmions, but often suffer from Joule heating, stochastic nucleation, or limited spatial selectivity \cite{fujita2017vortexbeams, Jia2015,Ber2018}. These constraints motivate the search for non-contact, ultrafast, and furthermore material-agnostic strategies for writing complex magnetic textures.

The question we address:  are there alternatives to the so far proven means of inducing topological states like skyrmion crystals. The answer we believe is in positive - one can use properly structured light beams to induce nontrivial magnetic textures.
Structured light offers precisely this combination of noncontact actuation and spatial programmability, with independently tunable polarization, azimuthal phase winding, and transverse mode structure across the optical spectrum from visible to telecom wavelengths~\cite{sliney2016light,All1992,gu2018gouy,wang2022orbital}.
Building on these optical concepts, analogous control fields have become feasible in the terahertz regime through recent advances in high field sources.
In parallel, advances in high-field terahertz sources have enabled phase-resolved transients with extreme field amplitudes, making it possible to drive magnetic dynamics through the magnetic component of the pulse on ultrafast timescales~\cite{sell2008phase,kampfrath2013resonant,Sederberg2020Tesla,kampfrath2011coherent,Kir2010}.

The concept of quantum printing has been demonstrated in a variety of platforms, including superconductors~\cite{Yeh2025a, Yeh2025b, Yerzhakov2024Kapitza, Yerzhakov2026pdwave}, Hall fluids~\cite{Car2026}, collective Higgs-mode dynamics~\cite{Mizushima2023, Kang2025, Kang2025KD}, and spin-orbit coupled electrons~\cite{Yamamoto2025}, establishing structured light as a universal and material-agnostic control knob for nonequilibrium quantum states.
A closely related ingredient is dynamical multiferroicity, where a time-dependent polarization generates a magnetization, a route recently demonstrated under intense circularly polarized terahertz driving in SrTiO$_3$~\cite{juras2017,Bas2024}.

Beyond these systems, magnetic materials provide a particularly rich arena for extending quantum printing into real space. Skyrmions have been identified across a wide range of magnetic materials and length scales, from atomic-scale textures in ultrathin films to mesoscopic skyrmion crystals, with their size and profile governed by competing magnetic energy scales~\cite{Muhl2009,Tok2021,Hei2011,Wu2021}. 
Moreover, skyrmions can be generated optically without OAM transfer~\cite{finazzi2013laser, je2018skyrmionbubble, buttner2021fluctuation, zhu2024ultrafast}, instead relying on topologically non-trivial lattice structures.

In this work we demonstrate quantum printing of noncollinear and skyrmionic spin textures using structured terahertz LG pulses. By coupling the vector optical magnetic field of light to a collinear ferromagnetic thin film, we show that the optical quantum numbers—spin angular momentum (SAM), orbital angular momentum (OAM), and radial order (RO)—act as independent and programmable control parameters for imprinting a broad hierarchy of magnetic textures. We use micromagnetic simulations to resolve the spatiotemporal evolution of magnetization and topological charge during the optical pulse, and demonstrate how optical symmetry and angular momentum selectively guide the formation of nontrivial spin configurations.

The paper is organized as follows. Section~\ref{sec:model} introduces the model and methods, we specify the effective Hamiltonian for the exchange coupled ferromagnetic film and the parameter regime considered; we then present the micromagnetic LLG simulations, the LG beam model, the definition and evaluation of the real space topological charge, and the quantum printing protocol that couples the optical field to the magnetization dynamics. Section~\ref{sec:results} presents the main printing outcomes, organized into skyrmion defects, domain wall-like defects, and higher order and radially structured textures. Section~\ref{sec:discussion} discusses the underlying symmetry selection and post pulse evolution of the printed states. Section~\ref{sec:conclusions} concludes with an outlook.

\section{Model and Methods}\label{sec:model}
\begin{figure}[ht!]
  \centering
  \includegraphics[width=0.85\columnwidth]{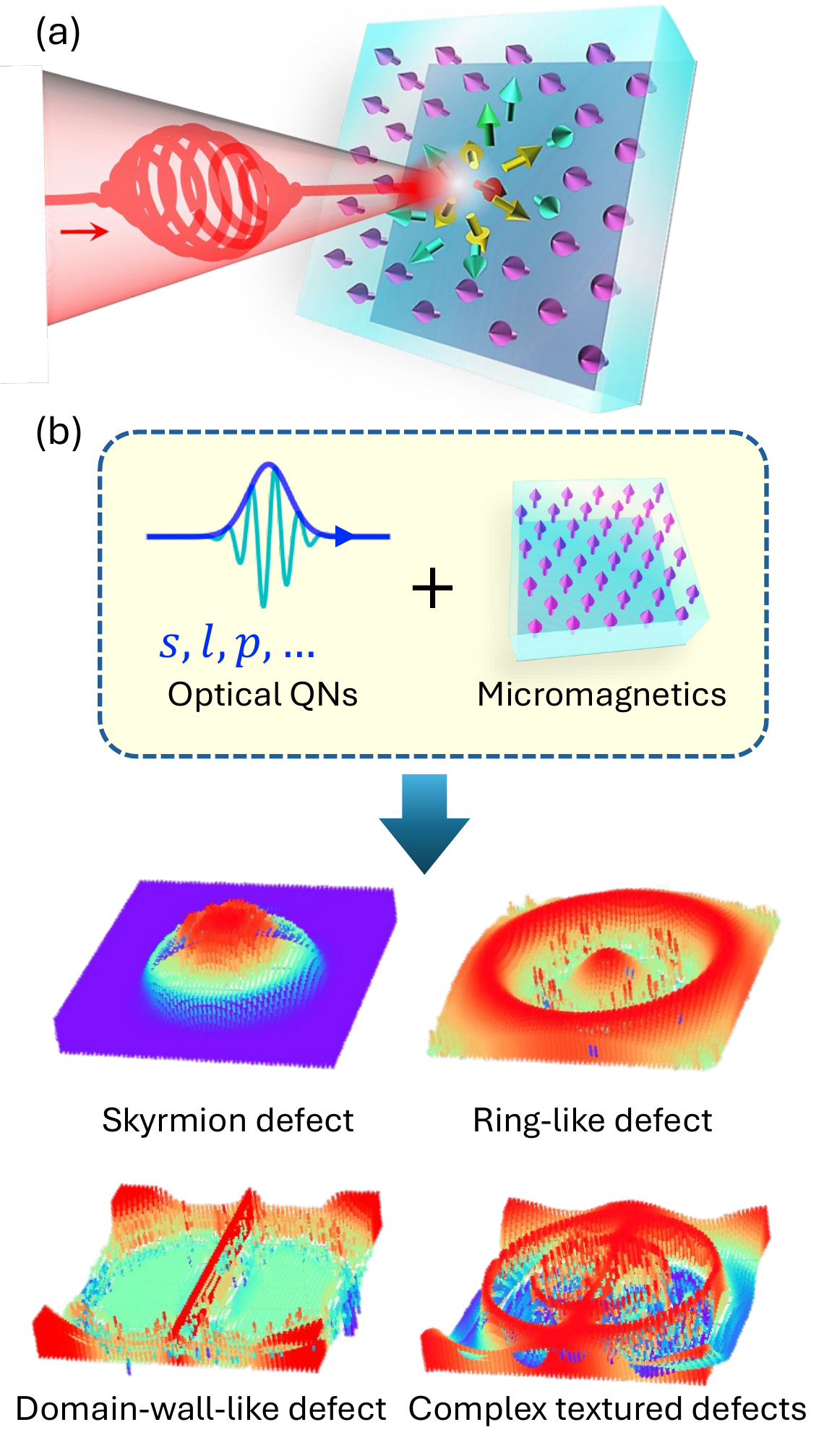}
  \caption{(Color online) 
    (a) Schematic of light-driven spin dynamics in a micromagnetic thin film, where a structured optical beam transfers angular momentum to the spins. 
    (b) Structured light carrying optical quantum numbers \((s ,\, \ell,\, p,\dots)\) imprints spatially varying non-collinear spin textures, generating topological non-trivial (skyrmion-like), ring-like, domain-wall-like, and other complex magnetic defects. 
  }
  \label{fig:fig1}
\end{figure}

Considering a collinear ferromagnetic (FM) thin film, a general Heisenberg spin Hamiltonian can be written as
\begin{equation}
  H = \sum_{\langle j,k \rangle} J_{j,k} {\bf m}_{j} \cdot {\bf m}_{k} - \sum_j \gamma_g {\bf B}_{\rm ext} \cdot {\mathbf m}_j,
  \label{eq:H}
\end{equation}
where $J_{j,k}$ is the nearest exchange coupling between site $j$ and its nearest neighbor site $k$ in the two-dimensional FM lattice, and the second term is the external magnetic field with gyromagnetic ratio $\gamma_g$. We only consider the collinear ferromagnetic system in this work.
Here, the static external magnetic field ${\bf B}_{\rm ext}$ is introduced for setting a well-defined initial magnetic state and providing a reference axis for the subsequent driven dynamics.

\textit{Diffraction limit of LG beams and non-collinear texture size.}
Atomistic spin dynamics (ASD) simulations are well suited for modelling magnetic systems at the nanometer scale, where atomic resolution is essential. However, extending ASD to micrometer-scale systems requires tracking millions of individual spin vectors, which becomes computationally heavy.
The spatial structure of light is fundamentally constrained by the diffraction limit, resulting in optical wavelengths typically ranging from hundreds of nanometers to several micrometers. For ultraviolet (UV) and infrared (IR) regimes, the corresponding oscillation frequencies are on the order of $\sim 10^{15}$ Hz and $\sim 10^{13}$ Hz, respectively~\cite{sliney2016light}.
By contrast, characteristic magnetic length scales—such as exchange lengths lie in the range of a few to tens of nanometers~\cite{ Hub1998, nagaosa2013topological}, while intrinsic spin dynamics occur at gigahertz to terahertz frequencies set by ferromagnetic resonance and exchange interactions~\cite{ Gil2004, Chu2015}. 
This large disparity in spatial and temporal scales leads to a significant mismatch between the electromagnetic field and the intrinsic spin  dynamics of the material.

To achieve a realistic correspondence between the light field and the spin system in both in wavelength and frequency, we consider a micrometer($\mu m$)-scale magnetic system in Eq.~\eqref{eq:H}, which couples naturally to the terahertz (THz) structured light. In this regime, the optical wavelength, frequency, and energy scale of the THz excitation are commensurate with the micromagnetic system size. Therefore, it is natural to employ a micromagnetic framework to simulate the interaction between structured light and the magnetic spin textures.

\subsection{LLG equation in micromagnetics }

We evolve the normalized magnetization field $\mathbf{m}(\mathbf{r},t)=\mathbf{M}/M_s$ by the LLG equation,
\begin{equation}
\partial_t \mathbf{m} = -\gamma\,\mathbf{m}\times \mathbf{B}_\mathrm{eff}
+ \alpha\,\mathbf{m}\times \partial_t \mathbf{m},
\label{eq:llg}
\end{equation}
with $\bf{m}$ being the unit magnetization vector that is described in micromagnetic dynamics, $\gamma$ the gyromagnetic ratio, and $\alpha$ the Gilbert damping parameter. In this work the effective field contains only the exchange contribution (considering $J_{j,k} = 5$ meV in Eq.~\eqref{eq:H} with Gilbert damping parameter $\alpha = 0.1$ of all simulations in this work), and the external field that serves as the optical drive in the effective field,
\begin{equation}
\mathbf{B}_\mathrm{eff} = \mathbf{B}_\mathrm{ex} + \mathbf{B}_\mathrm{ext}(t),
\label{eq:heff}
\end{equation}
where the micromagnetic exchange field $\mathbf{B}_\mathrm{ex} $, and $\mathbf{B}_\mathrm{ext}(t)$ can be a static external magnetic field $\mathbf{B}_\mathrm{ext}$, and/or also be a LG beam induced time-dependent magnetic field $\mathbf{B}_\mathrm{LG}(t)$. The latter is the key object in our work, which is introduced in the next subsection.
The continuum exchange energy is
\begin{equation}
E_\mathrm{ex}= \mathcal{A} \int (\nabla \mathbf{m})^2\, dV,
\end{equation}
which yields the micromagnetic exchange field
\begin{equation}
\mathbf{B}_\mathrm{ex}=\frac{2 \mathcal{A}}{\mu_0 M_s}\,\nabla^2 \mathbf{m}.
\label{eq:Hex_laplacian}
\end{equation}
Free boundaries imply the natural Neumann condition $\partial \mathbf{m}/\partial n=0$ on outer surfaces where only exchange is present with no external drive.
To connect with atomistic exchange, we use the small angle mapping that matches a Heisenberg spin spiral to the continuum energy. For atoms in cell $j$ with volume $V_j$ and pairwise exchange $J_{j,k}$ separated by projection $\Delta x_{j,k}$ along the spiral direction, the local exchange stiffness satisfies
\begin{equation}
\mathcal{A}_j = \sum_{k} \frac{J_{j,k}\,\Delta x_{j,k}^{2}}{2 V_j},
\end{equation}
and the macroscopic $\mathcal{A}$ enters Eq.~\eqref{eq:Hex_laplacian} as a volume average of the $\mathcal{A}_j$. For isotropic lattices, this expression reduces to the heuristic relation, $\mathcal{A}_j \propto \sum_{k} J_{j,k}/2$, which we use to set consistent orders of magnitude.
We solve Eqs.~\eqref{eq:llg}–\eqref{eq:Hex_laplacian} on a uniform Cartesian mesh with spacings $\Delta x,\Delta y,\Delta z$. The Laplacian is evaluated by centered second order finite differences with duplicated boundary values to enforce the Neumann boundary condition. Time integration uses a projected explicit scheme that preserves $|\mathbf{m}|=1$ cellwise. In practice, we employ a fourth order Runge-Kutta step in parameter scans. After every substep, the magnetization vector in each cell is renormalized to an unit length. Stability and accuracy are checked by halving the time step and refining the grid until the maximum torque norm and key observables converge within the stated tolerances.
For numerical stability, we integrate in dimensionless units using a field scale $B_0$ and a time unit $T_0=1/(\gamma B_0)$, i.e., the corresponding time scale $\sim 0.7$ ps, while the order of exchange coupling $J_{j,k} \sim 1$ meV~\cite{fujita2017vortexbeams}.
All simulations in this work are performed on a two dimensional ferromagnetic thin film discretized into an $80\times80$ square array of coarse grained micromagnetic cells, with grid spacing $\Delta x=\Delta y=10~\mu{\rm m}$, corresponding to a lateral size of $800~\mu{\rm m}\times800~\mu{\rm m}$.
The detailed space time profile of $\mathbf{B}_\mathrm{ext}(t) = {\bf B}_{LG}(t)$ for the optical LG drive is given in the upcoming Subsection~\ref{subsec:LG beam}. This choice of parameters is representative and is not restrictive. Detailed frequencies and length scales would depend on the specific properties of material, that is beyond the discussion of this protocol-focused study. 
\begin{figure*}[ht!]
  \centering
  \includegraphics[width=1.5\columnwidth]{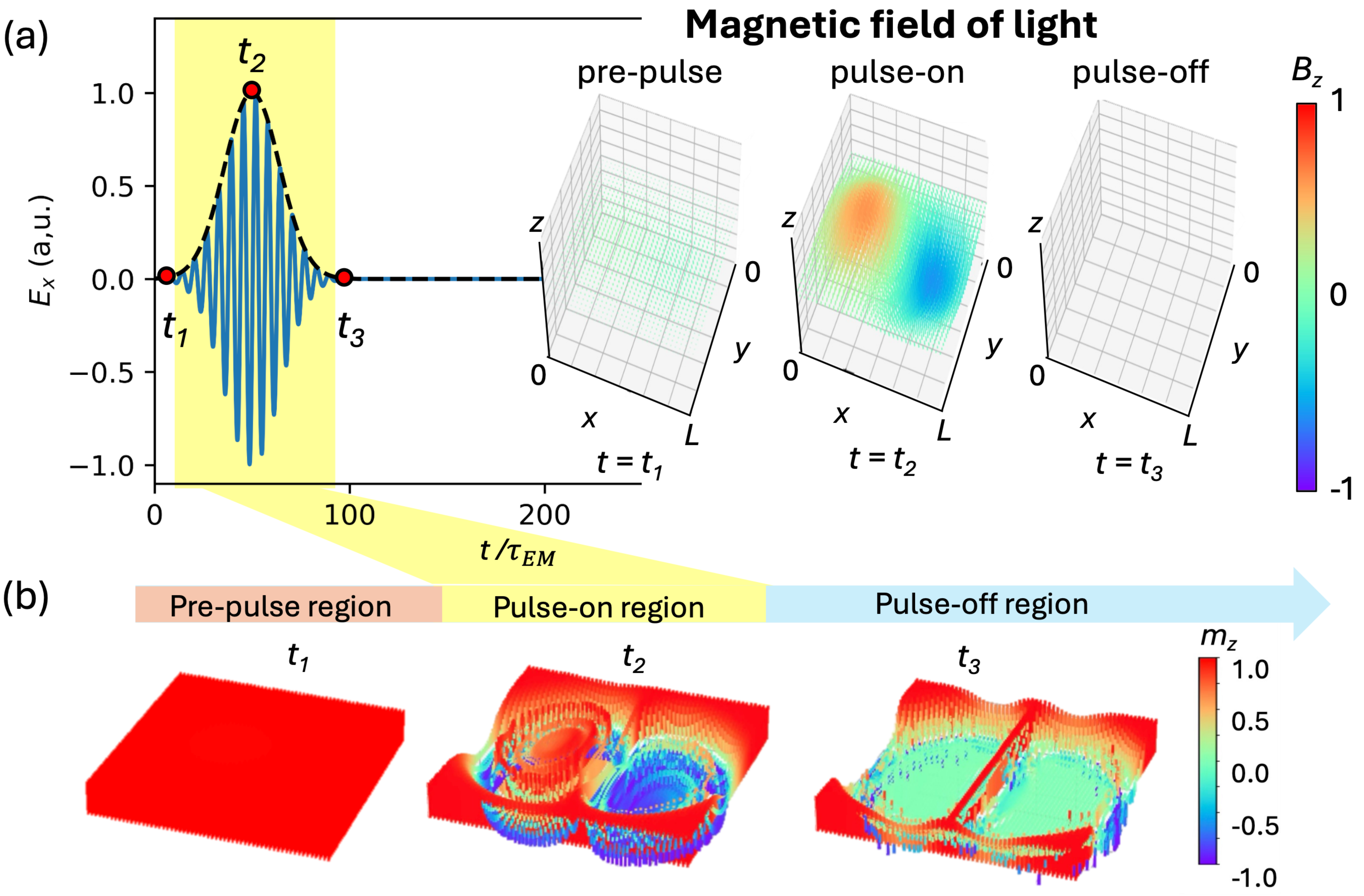}
  \caption{(Color online) 
    Demonstration of optical pulse timing and spin response in this work, using a linearly polarized optical pulse as a representative example.
    (a) Temporal profile of the electric field $E_x(t)$ used in the simulation. The representative times $t_1$–$t_3$ correspond to $t/\tau_{EM}=$ 10, 50 and 100, respectively. The shaded yellow region indicates the pulse-on interval. Right: snapshots of the out-of-plane magnetic field $B_z$ of the optical pulse at the pre-pulse ($t=t_1$), pulse-on ($t=t_2$), and pulse-off ($t=t_3$) stages.
    (b) Corresponding evolution of the spin texture at the same time frames: pre-pulse ($t_1$), pulse-on ($t_2$), and pulse-off ($t_3$). The linearly polarized light induces transient non-collinear spin structures that gradually relax after the pulse termination. 
    The spot size of the structure light is $2w_0 \sim 400$ $\mu m$, $J_{j,k}=5$ meV, and Gilbert damping parameter $\alpha$ is set to 0.1 for all simulations in this work.
}
  \label{fig:fig2}
\end{figure*}

\subsection{Laguerre-Gaussian beam}\label{subsec:LG beam}

Laguerre-Gaussian (LG) beam is a typical light structure carrying a topological charge, which exhibits not only the rich polarization in structures of electric field but various magnetic field structures. The LG transverse mode can be described by the Laguerre polynomial $L_p^{\ell}(f(r))$, where $p$ is the radial order (RO), and $\ell$ is the azimuthal index representing the quantum number of orbital angular momentum (OAM, $\ell\hbar$). The amplitude of the electric field of a LG mode is given by~\cite{wang2022orbital, gu2018gouy}
\begin{multline} 
\label{eq:LG_u}
u_{\ell,p}\left(r,\varphi,z\right)= E_0 C_{p\ell} \left[\frac{\sqrt2r}{w\left(z\right)}\right]^{\ell} \frac{w_0}{w\left(z\right)}
\\ L_p^{\ell} \left(\frac{2r^2}{w^2\left(z\right)}\right) \cdot e^{-i\phi_{p\ell}\left(z\right)+i\frac{k r^2}{2q\left(z\right)}+i \ell \varphi} ,
\end{multline}
The term $C_{p\ell}=\sqrt{\mathstrut 2 p!/\pi (p+|\ell|)!}$ is utilized to normalize the amplitude.
Along the parapagation direction $z$, the radius of the beam spot is equivalent to $w(z)=w_0 \sqrt{\mathstrut (z^2+z_R^2)/z_R^2}$ with the beam waist $w_0$ and the Rayleigh distance $z_R=k \omega_0^2/2$, where $k$ is the wavenumber, i.e., $k = 2 \pi/ \lambda_{EM}$, where the wavelength is $\lambda_{EM}=c \tau_{EM}$ and the period of light is $\tau_{EM}$. Additionally, $\phi_{p\ell}(z)=(2p+\abs{\ell}+1) \arctan{(z/z_R)}$ is the Gouy phase, $\exp{ikr^2/2q}$ is related to the radius of curvature with $q(z)=z-iz_R$, and the OAM term $\exp{i \ell\varphi}$ provides spiral wavefront via controlling temporal phase delay $\ell\varphi$ around the azimuthal angle $\varphi$.
In the focal plane, i.e., $z=0$, the radius of the spot is $w(z)=w_0$, the Gouy phase is 0, and the curvature term $\exp{i kr^2/2q(z)}$ becomes the Gaussian distribution $\exp{-r^2/w_0^2}$.
Throughout this work, the sample is assumed to be located in the focal plane.

Along $x$- and $y$-direction with different component of polarization, it allows to carry the SAM number $s$ with delay between polarization, also famous as circular polarization. 
To generate transient optical magnetic fields with amplitudes approaching the tesla scale, we model the excitation as an intense ultrafast THz pulse~\cite{Sederberg2020Tesla, kampfrath2013resonant,sell2008phase}, whose temporal carrier envelope is modulated by a Gaussian function, i.e. $E_G= E_0 \exp{-(t-t_0)^2/\tau_G^2}$, where $t_0$, $\tau_G$ denote the pulse center and pulse width of the Gaussian envelope, respectively.
Combining with the previously mentioned LG beam and Gaussian pulse, the electric field is formulated as
\begin{multline} 
\label{eq:LG_E}
\boldsymbol{E_{EM}}(\textbf{r}, t) = 
\\ E_G \cdot u_{\ell,p}(r,\varphi,z)\cdot e^{i(kz - \omega_{EM} t + \varphi_{t0})}
\begin{pmatrix}
\cos{\varphi_{xy}}  \\
 e^{-i \sigma} \sin{\varphi_{xy}} 
\end{pmatrix}
\begin{pmatrix}
{\hat{\mathbf{e}}}_{x} \\
{\hat{\mathbf{e}}}_{y}
\end{pmatrix} ,
\end{multline}
where $\varphi_{t0}$, $\varphi_{xy}$, and $\sigma$ provides the initial temporal phase ($t=0$), an offsets of azimuthal angle of $xy$-polarization, and number of SAM $\sigma\hbar$ $(\sigma=s \cdot \pi/2)$, respectively. For linear polarization (LP) and circular polarization (CP), $s$ are an even and odd number, respectively. For example, we use $s=0$ for LP, $s=1$ for left-handed CP, and $s=-1$ for right-handed CP. The notation LG$_{s,\ell,p}$ denotes the mode of the LG beam.

The magnetic field of light is estimated by $\mathbf{B} =\curl{\mathbf{A}}$ where $\mathbf{A}$ is vector potential. At focal point in free space, i.e. $z=0$,
supposed there is no spherical field induced $z$-componant of $\mathbf{E}$, the vector potential can be written as $\mathbf{A}=(A_x,A_y,0)=(E_x/i\omega_{EM},E_y/i\omega_{EM},0)$; therefore, the 3D magnetic field is formulated as follows,
\begin{align}
\label{eq:B}
B_x(\textbf{r}, t) 
&= Re \Big\{-\partial_z A_y(\textbf{r}, t) \Big\}
= -Re \Big\{E_y(\textbf{r}, t)/c \Big\} 
, \notag  \\
B_y(\textbf{r}, t) 
&= Re \Big\{\partial_z A_x(\textbf{r}, t) \Big\}
= Re \Big\{E_x(\textbf{r}, t)/c \Big\}
, \notag  \\
B_z(\textbf{r}, t) 
&= Re \Big\{\partial_x A_y(\textbf{r}, t) - \partial_y A_x(\textbf{r}, t) \Big\} \notag \\
&= Re \Big\{ \frac{1}{i\omega_{EM}} \left[ \partial_x E_y(\textbf{r}, t) - \partial_y E_x(\textbf{r}, t) \right] \Big\}
.
\end{align}
Accordingly, the size of beam along the $xy$-plane impacts the out-of-plane magnetic field, unlike the in-plane magnetic fields which is independent on beam size. Taking into account the Gaussian beam (LG$_{00}$ mode), amplitude of $B_z$ is proportional to $1/w_0^2$~\cite{Yeh2025a, Yeh2025b}.
We implement the different angular momentum numbers with a fixed electrical field $E_0=1.5 \times 10^{10}$ V/m, which is correspond to the peak value of $B=50$ T along $xy$-plane and frequency $\omega_{EM}=1$ THz~\cite{Sederberg2020Tesla, Martín-Domene2024}.
The parameters of Gaussian-shaped pulse $t_0/\tau_{EM}$ and $\tau_{G}/\tau_{EM}$ are 50 and 20, respectively.
The spot size $2w_0 \sim 400$ $\mu m$ of the structured light has been set the same for achieving various magnetic texture printing.  
The complete LG beam quantum printing protocol is described in~Section~\ref{subsec:protocol}, where the threshold values of the key parameters are addressed.

\subsection{Topological charge }
We characterize the non-collinear spin texture by its topological charge (so-called skyrmion number $N_{\rm sk}$) ~\cite{nagaosa2013topological, Berg1981}
\begin{eqnarray}
\mathcal{Q}(t) &=& \int_{r<R} q(\mathbf r,t)\, d^2 r, \nonumber \\
q(\mathbf r,t) &=&\frac{1}{4\pi}\,\mathbf m(\mathbf r,t)\cdot\bigl(\partial_x\mathbf m(\mathbf r,t)\times\partial_y\mathbf m(\mathbf r,t)\bigr).
\label{eq:Q_def}
\end{eqnarray}
For a smooth texture whose magnetization approaches a uniform direction on the boundary, \(\mathcal{Q}\) is a non-zero integer that counts non-trivial topology, i.e., how many times \(\mathbf m\) wraps the unit sphere. 
The integer values \(\mathcal{Q}=\pm 1,\pm 2,\ldots\) correspond to stable skyrmionic textures. When the integration region does not enclose a complete object, or when the boundary is not uniform, \(\mathcal{Q}\) can take fractional values, for example, a meron carries \(|\mathcal{Q}|=1/2\), and other fractions can occur at multi-domain junctions. 
For example, integrating over the entire film where the magnetization is uniform at the edge yields \(\mathcal{Q}=\pm 1\) for a single skyrmion. In contrast, integrating only over a disk that contains one vortex core but not the full texture returns \(\mathcal{Q}=\pm \tfrac{1}{2}\), which identifies a meron. As another example, if the disk cuts through a skyrmion, \(\mathcal{Q}(R)\) varies continuously with \(R\) and reaches an integer only when the integration region encloses the complete object and a uniform boundary.
We analysis some of the quantum printed textures by calculating the topological charge of the results shown below.

\subsection{LG beam quantum printing protocol to magnets}\label{subsec:protocol}

\begin{figure*}[ht!]
  \centering
  \includegraphics[width=1.5\columnwidth]{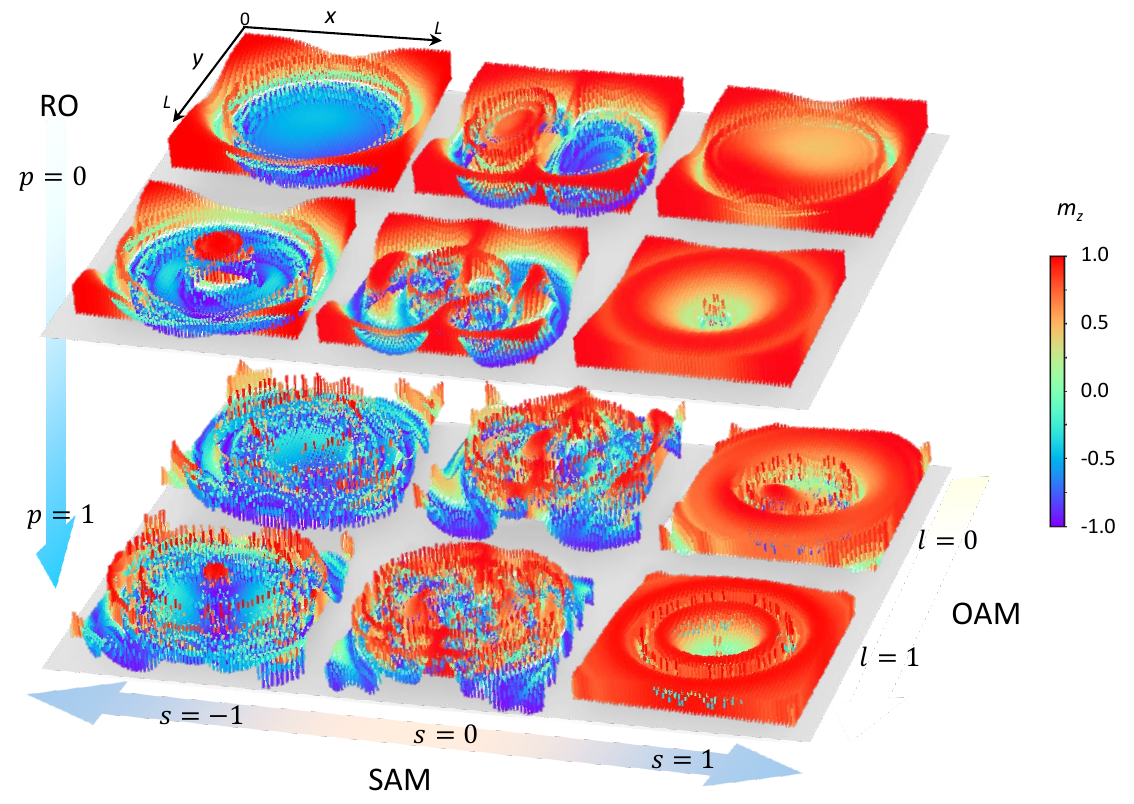}
  \caption{(Color online) 
 Spin textures induced by structured light with different optical quantum numbers ($s$,$\ell$,$p$). Snapshots near the pulse maximum at $t_2/\tau_{EM} = 50$ show characteristic magnetization patterns depending on the SAM ($s$), OAM ($\ell$), and RO ($p$) of the incident light.
 The spot size of the structure light is $2w_0 \sim 400$ $\mu m$, exchange coupling parameter $J_{j,k}=5$ meV, and Gilbert damping parameter $\alpha$ is set to 0.1 for all simulations in this work.
}
  \label{fig:fig3}
\end{figure*}

\begin{figure}[h!]
  \centering
  \includegraphics[width=1\columnwidth]{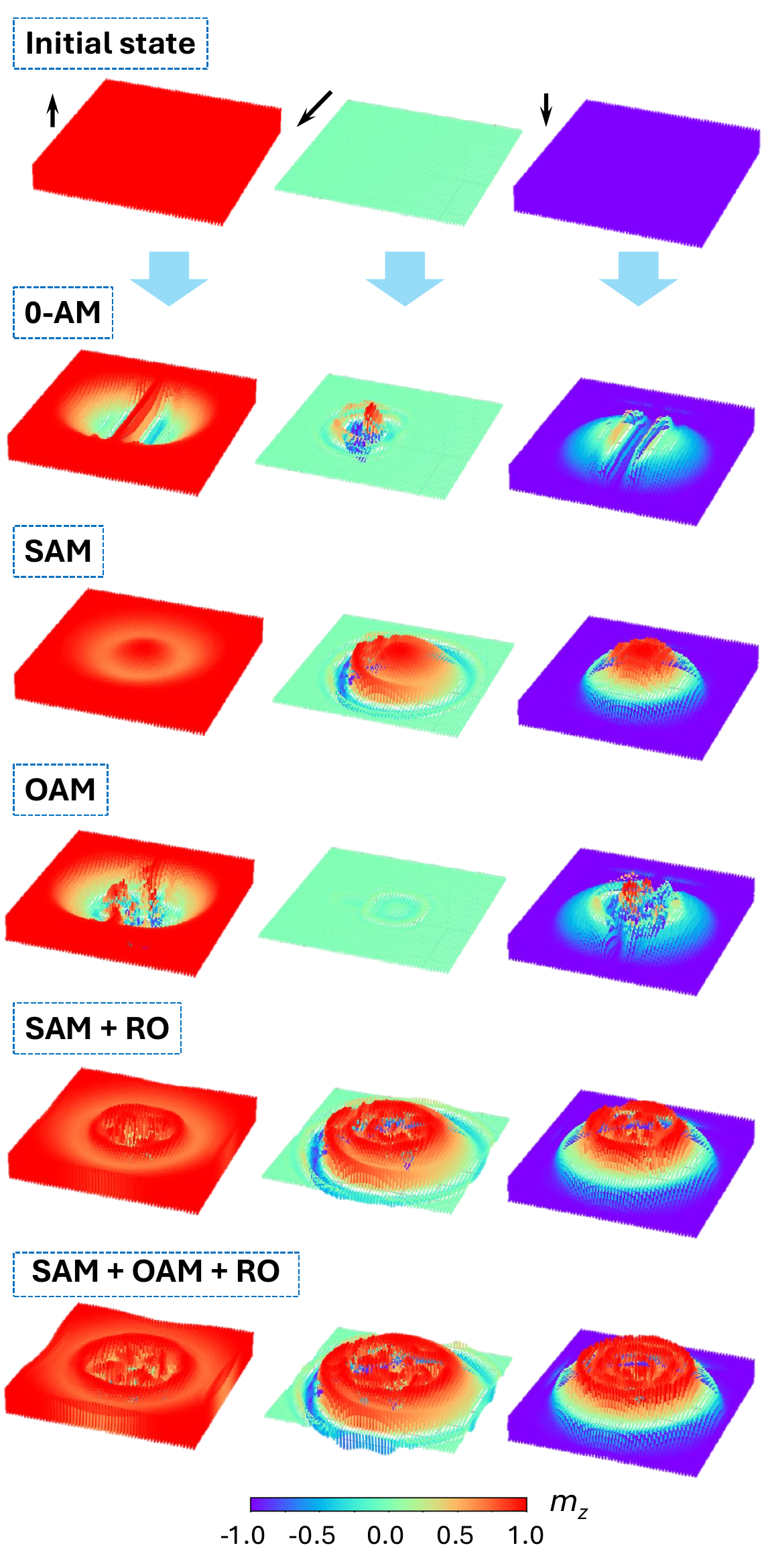}
  \caption{(Color online)
    Spin textures imprinted by structured light under different initial magnetic states and angular momentum numbers. 
    Top row: three initial states with uniform magnetization oriented up ($+z$), in-plane ($+x$), and down ($-z$). 
    Each column shows the resulting spin configuration snapshots at pulse-off time region at $t_3/\tau_{EM} = 200$ under illumination with different optical angular momentum components: 
    0-AM (no angular momentum), SAM ($s = 1$), OAM ($\ell = 1$), SAM+RO ($s,p = 1$), and SAM+OAM+RO ($s,\ell,p = 1$). 
    The color scale indicates the out-of-plane magnetization component $m_z$.}
  \label{fig:fig4}
\end{figure}

Figure~\ref{fig:fig1} summarizes the central simulation outcome of this work: structured light provides a highly flexible route for imprinting non-collinear magnetic textures in a ferromagnetic thin film. By tuning the optical quantum numbers $(s,\ell,p)$ associated with SAM, OAM, and radial order (RO), we obtain a rich variety of LG beam \textit{quantum printed} spin configurations. These include domain-wall-like defects, skyrmions, ring-like textures, and even defects with more complex textures. The resulting spin textures emerge directly from the spatial symmetry and angular momentum content of the optical field, demonstrating the principle of quantum printing: the controlled transfer of light’s internal quantum numbers into patterned magnetic order.

As shown in Fig.~\ref{fig:fig2}, the magnetization dynamics can be broadly divided into three temporal regimes. In the \emph{pre-pulse} stage ($t/\tau_{EM} < 10$), the spin configuration remains fully determined by the initial state, with negligible perturbation for assumed classical spin behavior. Once the laser pulse comes, the system enters the \emph{pulse-on} regime ($10  < t/\tau_{EM} < 100 $), where the optical magnetic field drives a rapid and strongly non-linear reorientation of spins. This period contains the dominant imprinting process, with the most dramatic change occurring near the pulse maximum at $t_2/\tau_{EM} = 50$. Finally, after the field subsides, the system enters the \emph{pulse-off} regime ($t/\tau_{EM} \geq 100$), where the spin texture undergoes partial relaxation, quenching into a metastable configuration that often differs substantially from the initial state. Unless otherwise noted, the analysis below focuses on the pulse-off imprinted textures, which encode the lasting effect of the structured light.

Our micromagnetic simulation considered a fixed LG beam profile, which is described in Section~\ref{subsec:LG beam}, for a $1$ THz structured light, a strong beam amplitude is required to print the quantum numbers in more detail, i.e., the magnetic field strength larger than $2 \pi\hbar\omega_{EM}/ g \mu_B \sim 30$ T.
Reducing the excitation frequency would proportionally lower the required magnetic-field amplitude, providing an alternative route toward experimentally accessible regimes.
The beam spot size is another parameter that affects the dynamics. Since the optical magnetic field is determined by the spatial variation of the vector potential, $\mathbf{B}=\nabla\times\mathbf{A}$, tighter focusing can enhance the local magnetic response and reduce the required electric-field amplitude.
From an experimental perspective, the beam spot size is highly tunable and can be reduced beyond the diffraction limit using, for example, scanning near-field optical microscopy (SNOM)~\cite{Guo2024Terahertz} and near field THz antenna~\cite{tuniz2023subwavelength} .
In addition, when the optical profile overlaps with the sample boundaries; while the micromagnetic simulation has considered an open boundary condition, as the magnetic samples are typical easy to be prepared larger than~{\rm mm} scale, this aspect is not considered here.

The optical drive is fully specified by three beam indices $(s,\ell,p)$, corresponding to spin angular momentum (helicity), orbital angular momentum (azimuthal phase winding), and radial order, respectively.
By scanning $s\in\{-1,0,1\}$, $\ell\in\{0,1\}$, and $p\in\{0,1\}$, we find that structured beams generate distinct classes of spin textures, as reflected by the spatial patterns of the out of plane magnetization component $m_z$ in Fig.~\ref{fig:fig3}. Importantly, these textures are not a point by point copy of the optical profile. The beam acts as a symmetry selected, spatially structured torque that seeds precession, while exchange redistributes the perturbation and Gilbert damping converts the driven dynamics into a relaxed, metastable configuration. As a result, the final patterns encode the optical symmetry and length scales set by $(s,\ell,p)$, but their detailed morphology is shaped by nonlinear relaxation rather than direct optical imprinting.

Fig.~\ref{fig:fig2}~(b) illustrates this imprinting process for the simplest case: a linearly polarized pulse corresponding to $(s,\ell,p) = (0,0,0)$.
In this case, no optical angular momentum is injected, yet the spatially localized terahertz pulse still induces a non-trivial redistribution of spins.
During the pulse-on stage, transient distortions appear at the excitation spot, eventually stabilizing into a domain-wall-like configuration in the pulse-off regime (e.g. at $t_3$). 
The complete simulation results shown in Figs.~\ref{fig:fig5-SKY} $-$ \ref{fig:fig8_left_right} are presented in the appended animations~\cite{qp-mm-github}. 
This behavior represents a purely optical quench—where the pulse locally suppresses the magnetization and triggers a driven-to-relaxed transition without requiring spin–orbit torque or topological optical modes.

The structured beam carries a well defined rotational symmetry and a handedness controlled by quantum numbers $s$ and $\ell$, and the magnetic response depends sensitively on how this symmetry is matched to the initial ferromagnetic state. This is illustrated in Fig.~\ref{fig:fig4}, where changing the initial magnetization direction between an in plane configuration and an out of plane configuration along $+{\mathbf z}$ or $-{\mathbf z}$ modifies both the symmetry and the efficiency of texture formation. An out of plane initial state preserves axial symmetry and favors azimuthally organized responses set by the optical winding.  In plane initial state introduces a preferred axis that competes with the beam symmetry and promotes anisotropic deformation and domain formation. Reversing the initial out of plane polarity selects the handedness of the subsequent precession direction and can favor or suppress specific chiral features. Taken together, Figs.~\ref{fig:fig3} and \ref{fig:fig4} show that the optically induced textures are jointly controlled by the beam symmetry and by the symmetry of the magnetic initial condition, with damping providing the route from a transient driven pattern to the observed final state.

\section{Quantum printing of non-collinear textures}\label{sec:results}

\subsection{Skyrmion defects}\label{subsec:skym}

A particularly instructive case is the generation of a full skyrmion with $N_{\rm sk} = |\mathcal{Q}|=1$ using a circularly polarized Gaussian beam with $(s,\ell,p)=(1,0,0)$ and an initially uniform state $\mathbf m_0=(0,0,-1)$, as evidenced by the emergence of a reversed core surrounded by a winding in plane texture in Figs.~\ref{fig:fig3} and \ref{fig:fig4}.
At first sight this outcome is counterintuitive because the applied optical magnetic field in this configuration has a vanishing out of plane component on axis, $B_{z}=0$, so there is no direct local Zeeman bias to flip the central spin.
The key is that the reversal is not a local copy of the optical profile but is governed by the LLG dynamics: rewriting Eq.~\eqref{eq:llg} as $\partial_t {\mathbf m}=-\gamma\,\mathbf m\times\bigl[\mathbf B_{\rm eff} -(\alpha/\gamma)\partial_t{\mathbf m}\bigr]$ we see how the transverse optical drive first induces coherent precession and tilts the spins in an annular region around the center, after which the damping term irreversibly rotates the magnetization toward the instantaneous effective field.
As soon as a radial gradient develops, exchange converts the surrounding tilt into a strong, self generated exchange field at the center through $\mathbf B_{\rm ex}\propto\nabla^2\mathbf m$, producing an effective out of plane bias even when the external field has $B_{z}=0$.
This feedback concentrates the relaxation into a localized reversal region, so the core can flip and lock into a skyrmion configuration whose topology is then sustained by the exchange stiffness.
In this sense, the skyrmion nucleation is damping assisted and exchange mediated: the beam seeds the symmetry and chirality, while the nonlinearity of the LLG dynamics provides the mechanism that translates a purely transverse drive into an out of plane core reversal.
\begin{figure}[ht!]
    \centering
    \includegraphics[width=0.95\linewidth]{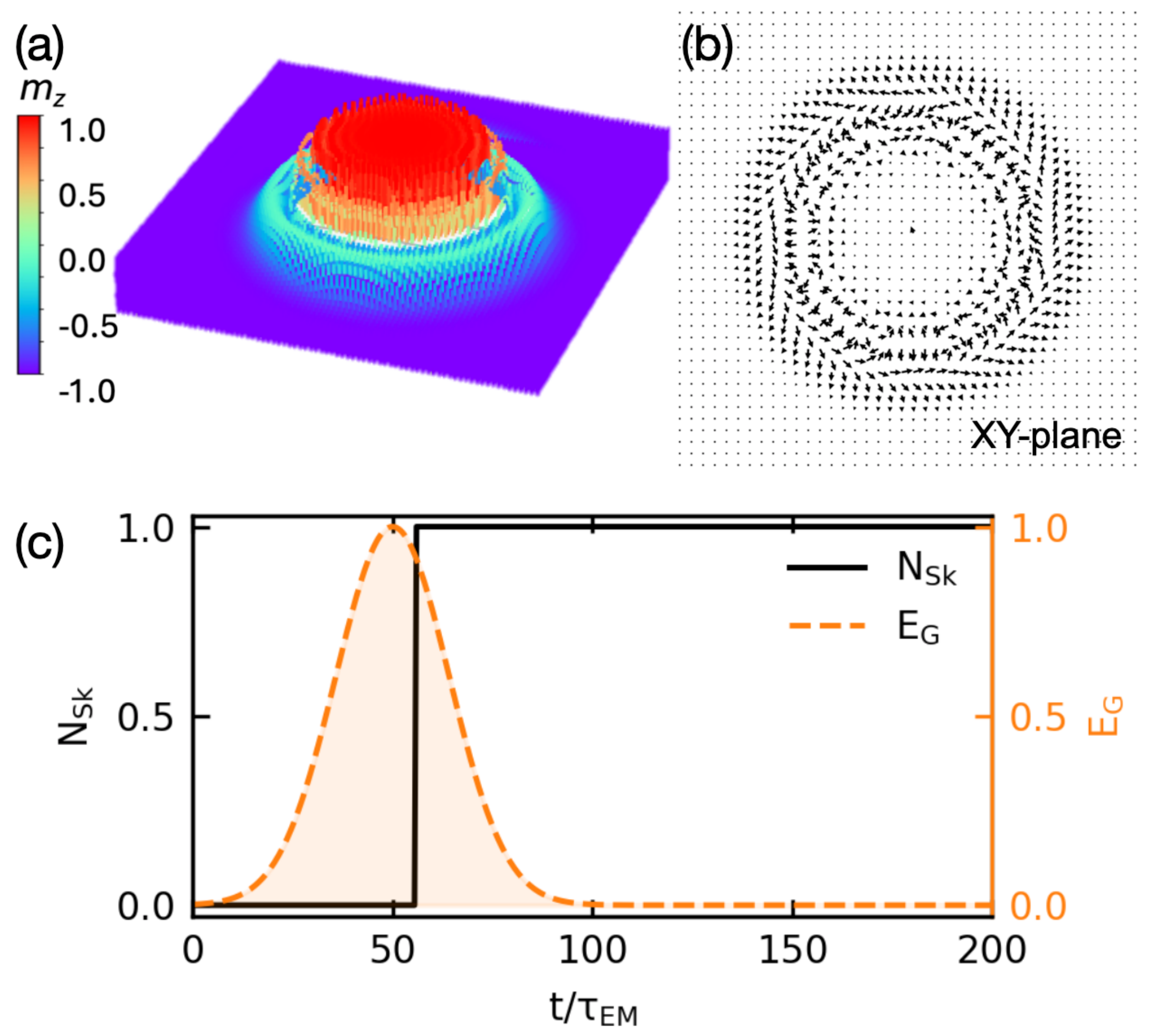}
    \caption{(Color online.)  Imprinted skyrmion in an exchange coupled ferromagnetic film. The optical mode is $(s,\ell,p)=(1,0,0)$ and the initial state $\mathbf m_0=(0,0,-1)$.
    (a) Three dimensional rendering of the magnetization with color indicating the out of plane component $m_z$, which is a snapshot after the pulse at $t_3/\tau_{EM} = 100$. A reversed core with $m_z\simeq +1$ is embedded in a background polarized with $m_z\simeq -1$.
    (b) In plane magnetization map $(m_x,m_y)$ in the film plane, showing a circulating texture around the core.
    (c) Time evolution of the skyrmion number $N_{\rm sk}$ (black, left axis) together with the normalized drive envelope $E_G$ (orange dashed, right axis). The topological charge switches from $N_{\rm sk}\approx 0$ to $N_{\rm sk}\approx 1$ during the pulse and remains pinned after the drive vanishes.}
    \label{fig:fig5-SKY}
\end{figure}

Figure~\ref{fig:fig5-SKY}~(c) highlights a  process of skyrmion printing.
We see a clear separation of time scales between writing and relaxation.
The transition to $N_{\rm sk}\simeq 1$ occurs within a short temporal window set by the pulse-on envelope, after which the topological charge remains quantized even as the LG beam field decays to zero, persisting for more than 50 pulse durations.
This persistence indicates that the system has entered a metastable basin in configuration space, so the subsequent dynamics is governed mainly by the topological protection and the slow energy dissipation into extended spin wave modes in the surrounding uniformly magnetized region.
As a result, the written texture behaves as a topological protected object on the simulation time window, which is advantageous for robust optical patterning in macroscopic samples.
In later Section~\ref{sec:discussion}, the slow energy dissipation is discussed as an overall phenomenon among all micromagnetic simulation outcomes in this work.

\subsection{Domain wall-like defects}\label{subsec:dw}

Domain-wall-like textures represent another class of imprinted spin structures in our simulations and typically arise under conditions where no net angular momentum is transferred from the light. As shown in Fig.~\ref{fig:fig2} and Fig.~\ref{fig:fig4}, such defects are most commonly generated using linearly polarized beams with $(s,\ell,p) = (0, 0, 0)$, where neither SAM nor OAM is injected into the system. In this regime, the light acts primarily as a localized quench: it transiently suppresses the magnetization amplitude near the beam center, creating a region of reduced order that partially recovers once the pulse is switched off.

Whether this quench evolves into a stable domain-wall texture depends on two ingredients. First, the system must possess sufficient exchange stiffness to support a coherent reversal front, rather than collapsing into disordered fluctuations. Second, the initial magnetization must be misaligned with the local field induced during the pulse, enabling a sign change in the out-of-plane component $m_z$ across the illuminated region. Under these conditions, the optical pulse seeds a topologically trivial but energetically long-lived configuration equivalent to a bent or curved Bloch wall.
\begin{figure}
    \centering
    \includegraphics[width=0.95\linewidth]{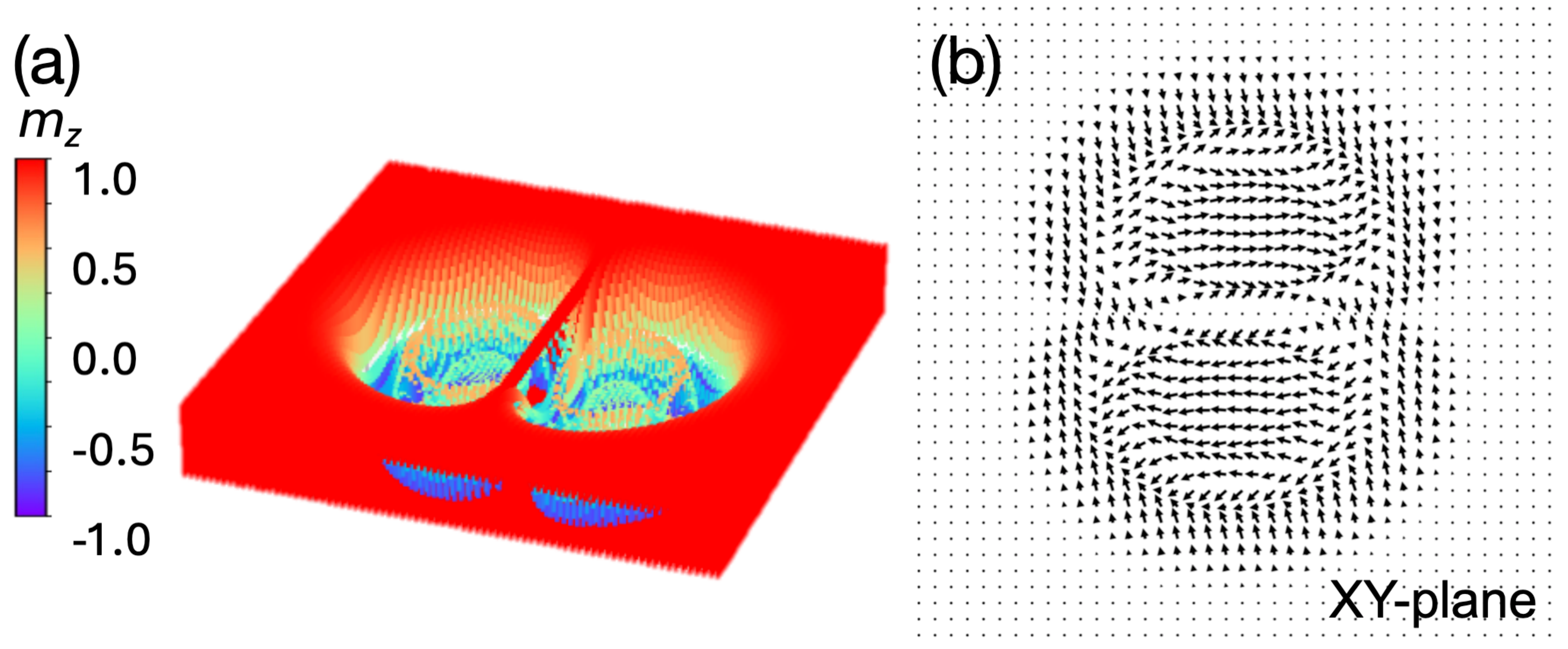}
    \caption{Imprinted domain wall-like  texture.
    The incident beam is linearly polarized light, corresponding to $(s,\ell,p)=(0,0,0)$, and the initial state of the magnetic system is $\mathbf m_0=(0,0,1)$. (a) Three dimensional rendering of the magnetization with color indicating the out of plane component $m_z$, which is a snapshot after the pulse at $t_3/\tau_{EM} = 100$. A narrow extended band of reduced $m_z$ forms a domain wall-like structure embedded in an otherwise uniformly polarized background, with pronounced end regions where the texture bends and localizes. (b) In plane magnetization map $(m_x,m_y)$ in the film plane. The arrows highlight a quasi-one dimensional rotation of the in plane component across the wall and a characteristic curling near the wall ends, indicating that the written object is an extended defect rather than an isolated core texture.}
    \label{fig:fig6-DW}
\end{figure}
Unlike skyrmions or merons, these domain-wall-like defects do not carry a quantized topological charge, and thus lack true topological protection. However, damping stabilizes them into metastable states that persist long after the pulse. This mechanism establishes the domain-wall imprinting process as a limiting case of quantum printing—where topology is not injected, but spatial contrast is robustly written by light-driven quench dynamics.
Beyond isolated skyrmionic cores, Fig.~\ref{fig:fig6-DW} demonstrates that structured light can also write extended line defects whose geometry is set by the competition between exchange stiffness and relaxation. The wall appears as a strip where the out of plane polarization is locally suppressed and the magnetization rotates through the film plane, consistent with an exchange dominated domain wall profile.
After the drive is removed, the subsequent evolution is governed mainly by curvature driven wall tension and the slow redistribution of exchange energy into long wavelength spin excitations.
In the macroscopic lateral sizes considered here, reaching the millimeter scale [Section~\ref{sec:discussion}], this relaxation becomes intrinsically sluggish because the energy stored along an extended wall must be transported over large distances before it can be dissipated efficiently.

\subsection{Higher-order and radially structured defects}\label{subsec:complex}

In addition to skyrmions, merons, and domain-wall textures, structured light with nonzero radial order ($p>0$) can generate even more complex magnetic configurations. As shown in Fig.~\ref{fig:fig4}, these modes introduce additional nodal rings in the optical field, which seed multiple vortex–antivortex pairs during the pulse.  When combined with finite OAM, the resulting textures can carry a total topological charge $|\mathcal{Q}| > 2$, particularly in cases where $|\ell - p| > 1$. In these regimes, the imprinting produces petal-like or multi-core patterns, reflecting the higher azimuthal winding encoded in the beam.

A notable feature of these higher-order textures is their spatially nonuniform relaxation. Spins at the beam center respond almost instantaneously to the light pulse, undergoing rapid reorientation and partial demagnetization. In contrast, spins outside the illuminated region respond only indirectly, through exchange-mediated coupling. This mismatch produces a hierarchy of timescales: fast local response inside the optical pulse duration, and slower collective reorganization in the surrounding region. The final imprinted pattern is therefore larger than the optical beam itself, extending outward as the system relaxes under damping.

These results demonstrate that complex LG beams (both $s, \ell \geq 1$, further with RO larger than one) act as programmable spatial masks, enabling quantum printing of complex textures, which are shown in the main text and summary in Supplementary Material~\cite{SI}. 
The combination of nodal structure and angular momentum opens a route to controlled multi-core and high-$\mathcal{Q}$ magnetic states, offering new possibilities for light-driven topology in spin systems.
\begin{figure}[ht!]
    \centering
    \includegraphics[width=0.95\linewidth]{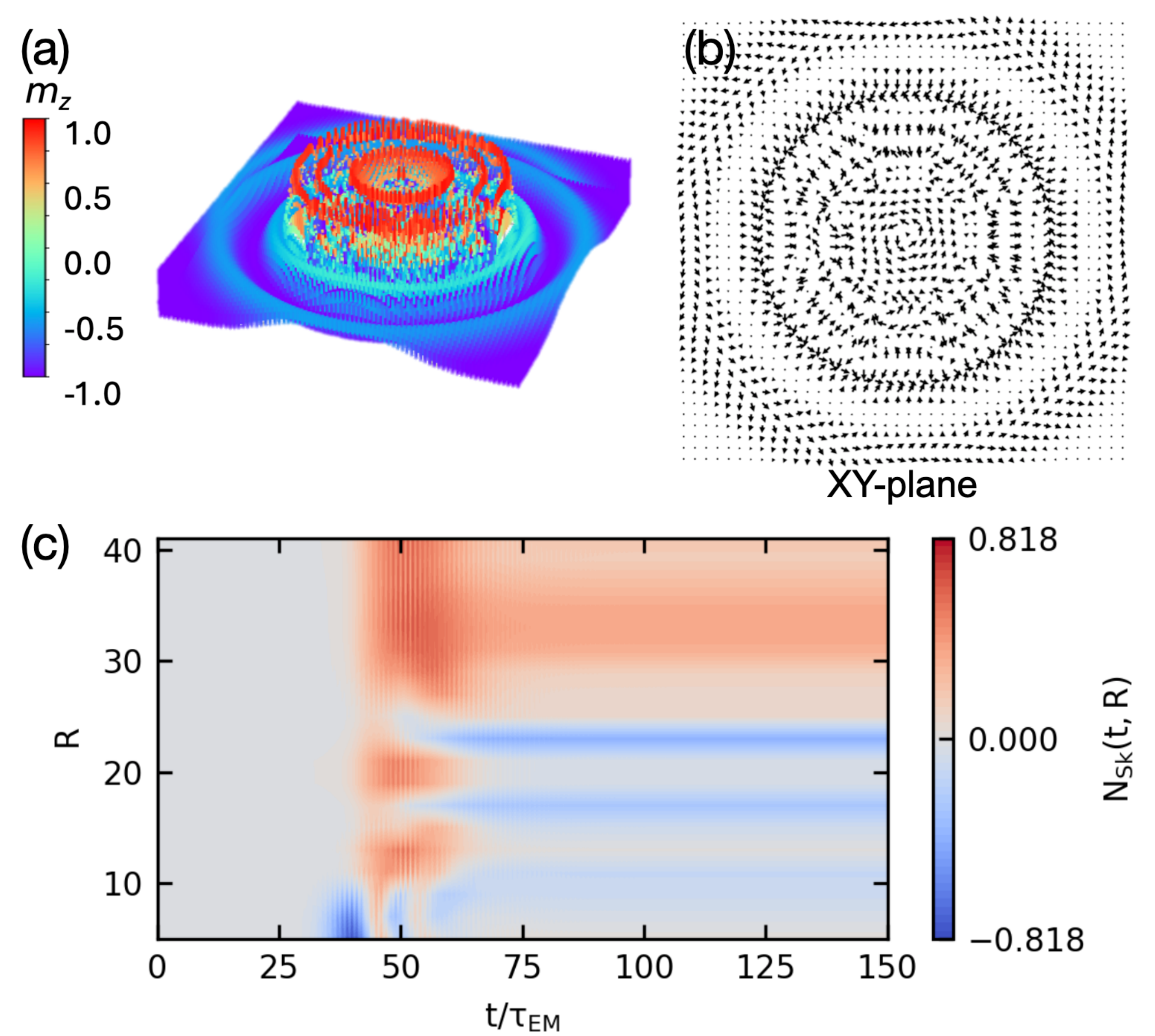}
    \caption{ Quantum printed complex spin ring for $(s,\ell,p)=(1,1,1)$ and initial state of magnetic system $m_0 = (0,0,-1)$.
(a) Three dimensional rendering of the magnetization with color indicating the out of plane component $m_z$, which is a snapshot after the pulse at $t_3/\tau_{EM} = 100$. The written state forms a concentric, multi ring structure with alternating regions of enhanced and suppressed $m_z$, indicating a hierarchical texture rather than a single core object.
(b) In plane magnetization map $(m_x,m_y)$ in the film plane. The arrows reveal a circulating pattern with strong azimuthal winding and radial modulation, consistent with a vortex like texture decorated by a ring node.
(c) Spatiotemporal contour map of the cumulative skyrmion number $N_{\rm sk}(t,R)$ as a function of time and integration radius $R$. During the pulse window the topological charge is redistributed across radii and subsequently settles into a persistent radial profile, demonstrating that the final texture carries multiple length scales encoded in the optical mode.}
    \label{fig:fig7-complex}
\end{figure}
Figure~\ref{fig:fig7-complex} demonstrates that moving from the single mode cases to $(s,\ell,p)=(1,1,1)$ enables controlled writing of qualitatively more complex textures, in which vorticity and radial structure coexist in a single metastable object. The ring morphology in $m_z$ and the corresponding in plane circulation indicate that the optical quantum numbers act as programmable handles that go beyond selecting a single skyrmion core and instead organize the magnetization into a multi-scale pattern. This is quantified in Fig.~\ref{fig:fig7-complex} (c): the contour map $N_{\rm sk}(t,R)$ shows that the texture does not acquire its topology uniformly in space, but through a radial build up in which distinct annular regions contribute with different signs and magnitudes before the profile stabilizes. The resulting plateau like behavior of $N_{\rm sk}$ over extended ranges of $R$ provides a compact diagnostic of a composite ring texture with internally redistributed topological charge.
In this sense, the $(s,\ell,p)$ triplet functions is a practical control set for quantum printing of spatially structured magnetic states.
\begin{figure}[ht!]
    \centering
    \includegraphics[width=0.92\linewidth]{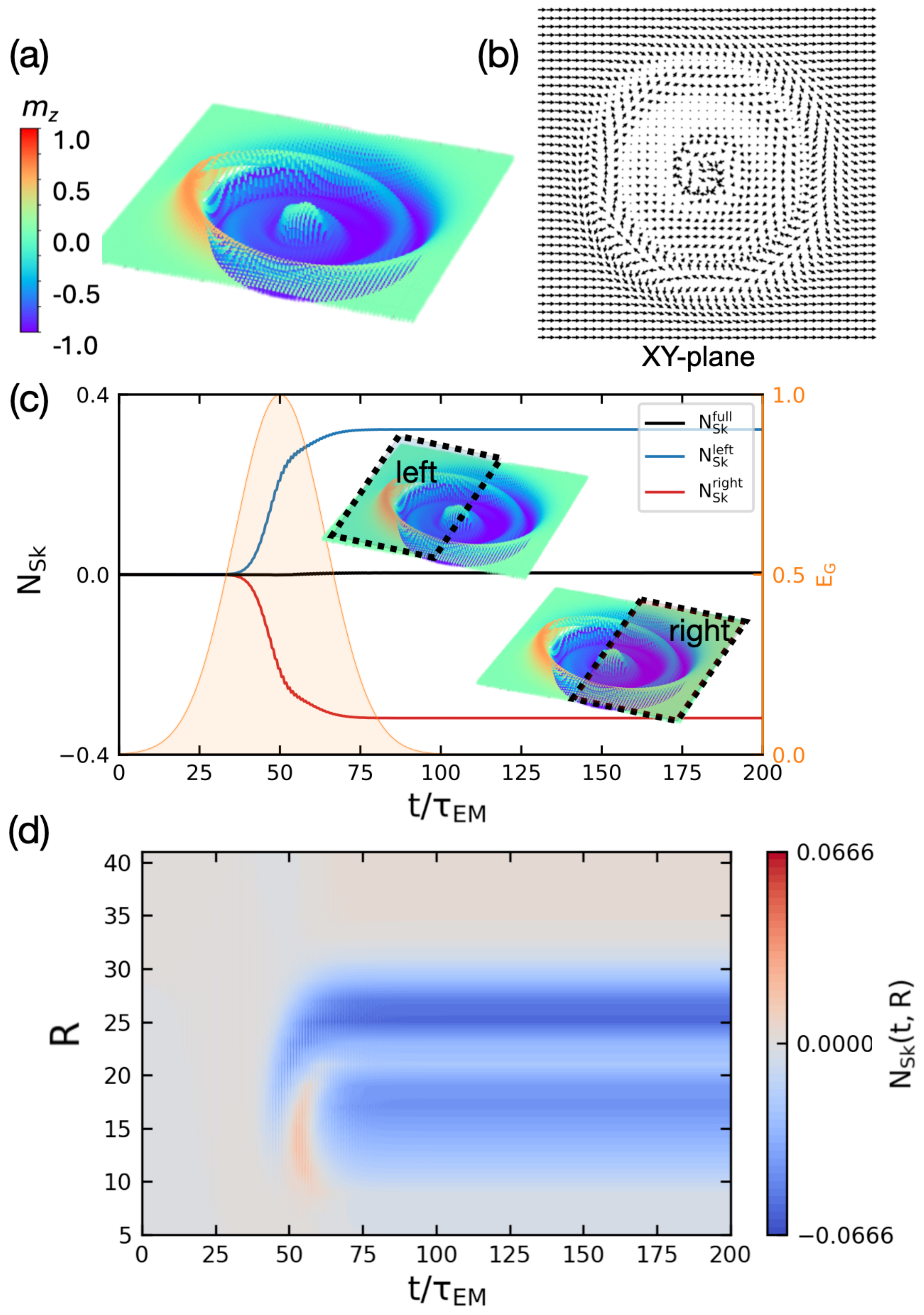}
    \caption{ Left right antisymmetric spin ring written by structured light.
The optical mode is $(s,\ell,p)=(-1,1,0)$ and the initial state is uniform and in plane, $\mathbf m_0=(1,0,0)$.
(a) Three dimensional rendering of the magnetization with color indicating the out of plane component $m_z$, which is a snapshot after the pulse at $t_3/\tau_{EM} = 100$.
The final state shows a ringlike modulation in $m_z$ that remains weak compared with the in plane rotation.
(b) In plane magnetization map $(m_x,m_y)$ in the film plane, revealing a complex ring texture whose handedness differs on the left and right sides of the pattern.
(c) Time traces of the skyrmion number evaluated over the full field of view, $N_{\rm sk}^{\rm full}$ (black), and over two complementary subregions, $N_{\rm sk}^{\rm left}$ (blue) and $N_{\rm sk}^{\rm right}$ (red), together with the normalized pulse envelope $E_G$ (shaded).
The partial charges acquire opposite signs while the full integral stays close to zero, indicating near cancellation between left and right contributions.
(d) Spatiotemporal contour map of the radius resolved charge $N_{\rm sk}(t,R)$, showing that the sign of the accumulated topology changes with the integration radius: a weak negative contribution builds up near the center (blue) and is compensated by a weak positive outer contribution (red) at larger $R$.
}
    \label{fig:fig8_left_right}
\end{figure}

Figure~\ref{fig:fig8_left_right} isolates a regime where the principal optical control is exerted on the in plane magnetization, while the net topological charge remains small. Starting from an in plane polarized ferromagnet, $\mathbf m_0=(1,0,0)$, the circular polarization and the azimuthal phase winding of the $(s,\ell,p)=(-1,1,0)$ beam do not translate into a single globally chiral texture. Instead, the driven response organizes into a left right antisymmetric ring in which the in plane circulation on one side is mirrored by an opposite circulation on the other side, as seen in Fig.~\ref{fig:fig8_left_right} (b). This cancellation is quantified by the partial integrals in Fig.~\ref{fig:fig8_left_right} (c): $N_{\rm sk}^{\rm left}$ and $N_{\rm sk}^{\rm right}$ separate during the pulse with opposite signs, yet $N_{\rm sk}^{\rm full}$ remains near zero, showing that the texture carries substantial local winding that is not captured by a single global integer invariant. The radius-resolved diagnostic in Fig.~\ref{fig:fig8_left_right} (d) further reveals a nontrivial radial redistribution of this winding: the accumulated charge changes sign as $R$ increases, with a weak inner contribution opposite to the outer contribution. Taken together, these observations establish that the same quantum printing protocol can be tuned to target predominantly planar textures, where the controllable outcome is the spatial arrangement and relative handedness of in plane rotations rather than a large net skyrmion number. This provides a practical route for mode selective engineering of in plane spin patterns, which is directly relevant for controlling phase and scattering of spin wave in extended ferromagnetic films.

\section{Discussion}\label{sec:discussion}

We presented the route to print magnetic structures in the collinear ferromagnets using structured light beams with optical quantum numbers (optical QNs) $(s,\ell,p) = (1,0,0)$.
The interesting result we find is that the imprinted magnetic textures can be topologically non-trivial: we demonstrate the ability to induce skyrmions in~Section~\ref{subsec:skym}.
We show that generation of the sykrmions with structured light can be achieved in the magnets with no DM interactions and with purely FM ground states with no frustration.
Hence we posit that the induced topological structure is inherited, i.e., imprinted by light structure.
Indeed for linearly polarized beams, we never observed the skyrmion formation.

Note that, in our micromagnetic simulation, $J_{j,k}\sim 1~{\rm meV}$ denotes the microscopic exchange coupling between neighboring atomistic spins. Each $10~\mu{\rm m}$ cell therefore represents a collective magnetization formed by a macroscopic number of atomic spins, so short wavelength fluctuations are averaged out inside the cell. After the pulse, the remaining texture varies only over mesoscopic to macroscopic length scales; the exchange torque is then controlled by the small curvature of the continuum field, $\sim \nabla^2\mathbf m$ in Eq.~\eqref{eq:Hex_laplacian}, rather than by the bare atomic exchange energy alone. This scale separation makes the pulse-off relaxation slow and allows the optically printed texture to remain metastable over the simulated time window.
Within the computational ability of our micromagnetic simulation, the light induced textures are persistent for a long time, i.e., the persistence time periods exceed 50 times of pulse duration.
That persistence makes these structures quasi-static, regardless of the topological properties of the imprinted magnetic textures.
These long persistence time might be useful for applications of light driven textures.  
 
The imprinted magnetic textures can also be topologically trivial, e.g., domain wall-like magnetic textures for the case of linearly polarized light, as shown in~Section~\ref{subsec:dw}.
We show that in the simplest case of linearly polarized light $(s,\ell,p) = (0,0,0)$, the pulse is able to induce long-lived magnetic textures through a purely local quench mechanism.
During the pulse-on stage, the intense THz field partially suppresses the magnetization amplitude in the illuminated region, effectively reducing the local exchange field and destabilizing the collinear ground state.
Once the pulse is switched off, the magnetization recovers, but not uniformly: spatial gradients in the recovery drive the formation of a domain-wall-like structure that persists into the pulse-off regime.

SAM- and OAM-mediated printing provides a route to topology-selective control by directly encoding phase winding and angular momentum into the light field.
Circular polarization ($s \neq 0$) or orbital angular momentum ($\ell \neq 0$) breaks the dynamical symmetry of the quench process by prescribing a preferred sense of rotation and vorticity.
We show two distinctive examples in~Section~\ref{subsec:complex}, and more examples can be found in our appended material~\cite{qp-mm-github}.
The role of SAM/OAM is to guide the reordering process by lifting degeneracies among competing configurations and stabilizing specific topological outcomes.
Recovery of magnetization is biased toward vortex configurations with a well-defined chirality or winding number, leading to deterministic imprinting of topological defects. 
SAM/OAM printing may be viewed as a controlled symmetry-breaking extension of optical quench imprinting, in which topology is not merely emergent but selectively enforced by the structure of the light field. 
We also want to point the differences and synergies  of the proposed mechanism to the thermal quench approach.
Even in the absence of SAM and OAM, linearly polarized beam can induce topological structures through quench, as was observed earlier~\cite{fu2018optical,Fan2023KZ}.
The quench process differs fundamentally from angular-momentum-mediated printing.
No phase winding is injected, and no topological charge is imposed by the optical field.
Instead, the system evolves through a dynamical symmetry breaking via interactions: the light first erases order, then the spin system reorders in a manner constrained by exchange, damping, and the spatial footprint of the pulse.

Optical quench imprinting with unstructured light represents topologically trivial limit of quantum printing: no phase winding or topological charge is directly imposed by the light field itself.
Nevertheless, this does not preclude the subsequent emergence of topological defects.
As an example of a non-topological quench dynamics, we point to  optical  drives the system across a critical point, where the recovery of magnetic order can proceed via a Kibble–Zurek–type mechanism~\cite{Fan2023KZ}.
Ponderomotive energy transfer leads to the spontaneous formation of vortex–antivortex textures even for linearly polarized excitation. Such quench-induced vortices have been directly observed experimentally~\cite{fu2018optical}. 
Interestingly, this mechanism coexists with SAM/OAM-based imprinting in more complex beam configurations.
Even when $s$ or $\ell$ is non-zero, a portion of the imprinted structure can be traced to this quench-and-recovery process, especially near the beam center where magnetization suppression is strongest.
Thus, optical quenching should be understood as a universal baseline mechanism of light-driven spin patterning upon which topology and winding can be selectively added.

Importantly, this angular-momentum-mediated printing does not replace the quench mechanism but acts on top of it.
The optical pulse still transiently suppresses the magnetization amplitude, providing the nonequilibrium pathway through which new textures can form.
From this perspective, optical quenching defines the nonequilibrium landscape of accessible spin textures, while SAM and OAM provide handles to sculpt that landscape by imposing phase and angular-momentum constraints.
Quantum printing thus continuously interpolates between the emergent and enforced topology, depending on the structure of the driving field.

\section{Conclusions}\label{sec:conclusions}
In this work, we established a structured-light \emph{quantum printing} route to generate and control noncollinear and topological spin textures in a collinear ferromagnetic thin film driven by terahertz LG pulses. By tuning the optical quantum numbers (SAM, OAM, and RO), our simulations demonstrate a broad hierarchy of imprinted patterns, ranging from domain-wall-like defects to skyrmion-like and multi-ring skyrmionic textures, with the topology and geometry dictated by the symmetry and phase structure of the incident field.

A key message is that these results are obtained within a deliberately minimal magnetic model: the only intrinsic interaction retained is the isotropic exchange coupling. In particular, no Dzyaloshinskii--Moriya interaction or other built-in chiral terms are required for the appearance of the printed textures in our simulations. The spin structures therefore do not originate from material-specific chirality, but from the externally imposed spatio-temporal structure of the optical magnetic field, which injects symmetry and angular-momentum constraints directly into the magnetization dynamics. This highlights quantum printing as a materials-agnostic principle for accessing complex non-collinear textures, even in otherwise conventional exchange-dominated ferromagnets \cite{qp2025}.

More broadly, our results clarify the separation between a universal nonequilibrium pathway and topology-selective control. The optical drive provides a transient dynamical window in which the collinear order is reconfigured, while the structured-light degrees of freedom bias the reordering process toward specific vorticities and defect geometries. In this sense, pulse dynamics provides for transiently suppression of the collinear order, whereas SAM/OAM and beam structure act as programmable field that selects and stabilizes particular topological textures during and after the pulse. The described process is qualitatively different from the pondermotive energy induced destruction of the order.  While the ponderomotive energy transfer is present in any pulse experiment, 
the proposed quantum-printing mechanism is not primarily governed by pulse-induced heating.

The present study provides a proof of principle foundation for magnetic quantum printing and suggests several natural extensions.
Incorporating anisotropy, dipolar fields, interfacial interactions, temperature, and realistic boundaries will allow quantitative predictions of robustness and lifetimes in specific materials.
Experimental implementations will require confined, high field structured terahertz sources, for example near field probes, nanoantenna arrays, or metasurface based emitters, to write magnetic textures at controllable length scales~\cite{Sederberg2020Tesla,Guo2024Terahertz,tuniz2023subwavelength,Niu2026Optica}.
Our findings therefore identify structured terahertz light as a route to optically programmable non-collinear and skyrmionic textures, and as a platform for developing future operator level descriptions of light to matter quantum transduction.

\begin{acknowledgments}
All authors acknowledge the insightful discussion with Deamo Kang, and fruitful discussion with Jared Cole, Hennadii Yerzhakov, Xiang Li, Takahiro Morimoto, Naoto Nagaosa, Xiuzhen Yu, Chih-Wei Luo.
Y.Liu is supported by  ERC No.2018-SYG 810451 HERO at Nordita.
T.T. Yeh. and A.V.B. are supported by the U.S. Department of Energy, Office of Science, Office of Basic Energy Sciences under Award No. DE-SC0025580 (concept, writing, editing)  and the Knut and Alice Wallenberg Foundation KAW 2019.0068. 
\end{acknowledgments}

\paragraph*{Data availability.}
Simulation data sets are available from the corresponding author on reasonable request. Software and the movies of the simulated dynamics with LG beam and LLG protocol are available openly in the Github repository at \url{https://github.com/yuefeiliuphys/qp-mm} .


\begin{thebibliography}{99}

\bibitem{nagaosa2013topological}
N.~Nagaosa and Y.~Tokura,
Topological Properties and Dynamics of Magnetic Skyrmions,
Nat.\ Nanotechnol.\ \textbf{8}, 899 (2013).

\bibitem{fert2017magnetic}
A.~Fert, N.~Reyren, and V.~Cros,
Magnetic Skyrmions: Advances in Physics and Potential Applications,
Nat.\ Rev.\ Mater.\ \textbf{2}, 17031 (2017).

\bibitem{qp2025}
G.~Aeppli, A.~V.~Balatsky, S.~Bonetti, G.~Cardoso, S.~Raghu, E.~Sylju{\aa}sen,
T.-T.~Yeh, S.-Z.~Lin, Y.~Liu, J.~Weissenrieder, and P.~J.~Wong,
Quantum Printing,
arXiv:2509.16792 (2025).

\bibitem{Bog1994}
A.~N.~Bogdanov and A.~Hubert,
Thermodynamically Stable Magnetic Vortex States in Magnetic Crystals,
J.\ Magn.\ Magn.\ Mater.\ \textbf{138}, 255 (1994).

\bibitem{Ros2006}
U.~K.~R\"ossler, A.~N.~Bogdanov, and C.~Pfleiderer,
Spontaneous Skyrmion Ground States in Magnetic Metals,
Nature \textbf{442}, 797 (2006).

\bibitem{yu2018transformation}
X.~Z.~Yu, W.~Koshibae, Y.~Tokunaga, K.~Shibata, Y.~Taguchi, N.~Nagaosa, and Y.~Tokura,
Transformation of Topological Spin Textures between Skyrmion and Meron in a Chiral Magnet,
Nature \textbf{564}, 95 (2018).

\bibitem{fujita2017encodingoam}
H.~Fujita and M.~Sato,
Encoding orbital angular momentum of light in magnets,
Phys.\ Rev.\ B \textbf{96}, 060407(R) (2017).

\bibitem{gao2024prl}
L. Gao, S. Prokhorenko, Y. Nahas, L. Bellaiche,
Dynamical control of topology in polar skyrmions via twisted light,
Phys.\ Rev.\ Lett.\ \textbf{132}, 026902 (2024).

\bibitem{shirato2025}
T.~Shirato, R.~Yambe, and S.~Hayami,
Bimeron Crystals by a Linearly Polarized AC Electric Field in Frustrated Magnets,
J.\ Phys.\ Soc.\ Jpn.\ \textbf{94}, 063601 (2025).

\bibitem{laxamana2026}
R.~J.~Laxamana and S.~Hayami,
Skyrmion generation via Laguerre-Gaussian beam irradiation in frustrated magnets,
arXiv:2603.03773 (2026).


\bibitem{fujita2017vortexbeams}
H.~Fujita and M.~Sato,
Ultrafast generation of skyrmionic defects with vortex beams,
Phys.\ Rev.\ B \textbf{95}, 054421 (2017).


\bibitem{Jia2015}
W.~Jiang, P.~Upadhyaya, W.~Zhang, G.~Yu, M.~B.~Jungfleisch, F.~Y.~Fradin, J.~E.~Pearson,
Y.~Tserkovnyak, K.~L.~Wang, O.~Heinonen, S.~G.~E.~te~Velthuis, and A.~Hoffmann,
Blowing Magnetic Skyrmion Bubbles,
Science \textbf{349}, 283 (2015).

\bibitem{Ber2018}
G.~Berruto, I.~Madan, Y.~Murooka, G.~M.~Vanacore, E.~Pomarico, J.~Rajeswari, R.~Lamb, P.~Huang,
A.~J.~Kruchkov, Y.~Togawa, T.~LaGrange, D.~McGrouther, H.~M.~R{\o}nnow, and F.~Carbone,
Laser-Induced Skyrmion Writing and Erasing in an Ultrafast Cryo-Lorentz Transmission Electron Microscope,
Phys.\ Rev.\ Lett.\ \textbf{120}, 117201 (2018).



\bibitem{All1992}
L.~Allen, M.~W.~Beijersbergen, R.~J.~C.~Spreeuw, and J.~P.~Woerdman,
Orbital Angular Momentum of Light and the Transformation of Laguerre--Gaussian Laser Modes,
Phys.\ Rev.\ A \textbf{45}, 8185 (1992).

\bibitem{gu2018gouy}
X.~Gu, M.~Krenn, M.~Erhard, and A.~Zeilinger,
Gouy Phase Radial Mode Sorter for Light: Concepts and Experiments,
Phys.\ Rev.\ Lett.\ \textbf{120}, 103601 (2018).

\bibitem{wang2022orbital}
J.~Wang, J.~Liu, S.~Li, Y.~Zhao, J.~Du, and L.~Zhu,
Orbital Angular Momentum and Beyond in Free-Space Optical Communications,
Nanophotonics \textbf{11}, 645 (2022).

\bibitem{sliney2016light}
D.~H.~Sliney,
What Is Light? The Visible Spectrum and Beyond,
Eye \textbf{30}, 222 (2016).

\bibitem{Kir2010}
A.~Kirilyuk, A.~V.~Kimel, and T.~Rasing,
Ultrafast Optical Manipulation of Magnetic Order,
Rev.\ Mod.\ Phys.\ \textbf{82}, 2731 (2010).

\bibitem{kampfrath2011coherent}
T.~Kampfrath, A.~Sell, G.~Klatt, A.~Pashkin, S.~M\"ahrlein, T.~Dekorsy, M.~Wolf,
M.~Fiebig, A.~Leitenstorfer, and R.~Huber,
Coherent Terahertz Control of Antiferromagnetic Spin Waves,
Nat.\ Photonics \textbf{5}, 31 (2011).

\bibitem{kampfrath2013resonant}
T.~Kampfrath, K.~Tanaka, and K.~A.~Nelson,
Resonant and Nonresonant Control over Matter and Light by Intense Terahertz Transients,
Nat.\ Photonics \textbf{7}, 680 (2013).

\bibitem{sell2008phase}
A.~Sell, A.~Leitenstorfer, and R.~Huber,
Phase-Locked Generation and Field-Resolved Detection of Widely Tunable Terahertz Pulses with Amplitudes Exceeding 100~MV/cm,
Opt.\ Lett.\ \textbf{33}, 2767 (2008).

\bibitem{Sederberg2020Tesla}
S.~Sederberg, F.~Kong, and P.~B.~Corkum,
Tesla-Scale Terahertz Magnetic Impulses,
Phys.\ Rev.\ X \textbf{10}, 011063 (2020).

\bibitem{Martín-Domene2024}
Sergio Martín-Domene, Luis Sánchez-Tejerina, Rodrigo Martín-Hernández, Carlos Hernández-García,
Generation of intense, polarization-controlled magnetic fields with non-paraxial structured laser beams,
Appl.\ Phys. \ Lett. {\bf 124}, 211101 (2024).

\bibitem{Yeh2025a}
T.-T.~Yeh, H.~Yerzhakov, L.~Bishop-Van~Horn, S.~Raghu, and A.~V.~Balatsky,
Quantum Printing and Induced Vorticity in Superconductors I: Linearly Polarized Light,
Phys.\ Rev.\ Research \textbf{7}, 043111 (2025).

\bibitem{Yeh2025b}
T.-T.~Yeh, H.~Yerzhakov, L.~Bishop-Van~Horn, S.~Raghu, and A.~V.~Balatsky,
Quantum Printing and Induced Vorticity in Superconductors II: Laguerre--Gaussian Beam,
Phys.\ Rev.\ Research \textbf{7}, 043112 (2025).

\bibitem{Yerzhakov2024Kapitza}
H.~Yerzhakov, T.-T.~Yeh, and A.~V.~Balatsky,
Induction of Orbital Currents and Kapitza Stabilization in Superconducting Circuits with Laguerre--Gaussian Microwave Beams,
Phys.\ Rev.\ B \textbf{110}, 144519 (2024).

\bibitem{Yerzhakov2026pdwave}
H.~Yerzhakov and A.~V.~Balatsky,
Induction of $p$-Wave and $d$-Wave Order Parameters in $s$-Wave Superconductors with Light Pulses,
arXiv:2602.09391 (2026).

\bibitem{Car2026}
G.~Cardoso, E.~Sylju{\aa}sen, and A.~V.~Balatsky,
Orbital Inverse Faraday and Cotton--Mouton Effects in Hall Fluids,
Phys.\ Rev.\ Lett.\ \textbf{136}, 016502 (2026).


\bibitem{Mizushima2023}
T.~Mizushima and M~.Sato,
Imprinting spiral Higgs waves onto superconductors with vortex beams,
Phys. Rev. Research {\bf 5}, L042004 (2023). 

\bibitem{Kang2025}
D.~Kang, S.~Kitamura, and T.~Morimoto,
Quantum Optical Spanner: Twisting Superconductors with Vortex Beam via Higgs Mode,
Phys. Rev. B {\bf 112}, 214315 (2025).

\bibitem{Kang2025KD}
D.~Kang, T.-T.~Yeh, T.~Morimoto, and A.~V.~Balatsky,
Kapitza--Dirac Interference of Higgs Waves in Superconductors,
arXiv:2511.10954 (2025).

\bibitem[]{Yamamoto2025}
S.~Yamamoto, M.~Sato, S.~Fujimoto, and T.~Mizushima, 
Optical vortex pulse induced nonequilibrium spin textures in spin-orbit coupled electrons,
Phys. Rev. B {\bf 112}, 205306 (2025).

\bibitem{juras2017}
D.~M.~Juraschek, M.~Fechner, A.~V.~Balatsky, and N.~A.~Spaldin,
Dynamical multiferroicity,
Phys.\ Rev.\ Materials \textbf{1}, 014401 (2017).

\bibitem{Bas2024}
M.~Basini, M.~Pancaldi, B.~Wehinger, M.~Udina, V.~Unikandanunni, T.~Tadano, M.~C.~Hoffmann, A.~V.~Balatsky, and S.~Bonetti,
Terahertz Electric-Field-Driven Dynamical Multiferroicity in SrTiO$_3$,
Nature \textbf{628}, 534 (2024).


\bibitem{Muhl2009}
S.~M\"uhlbauer, B.~Binz, F.~Jonietz, C.~Pfleiderer, A.~Rosch, A.~Neubauer, R.~Georgii, and P.~B\"oni,
Skyrmion Lattice in a Chiral Magnet,
Science \textbf{323}, 915 (2009).

\bibitem{Tok2021}
Y.~Tokura and N.~Kanazawa,
Magnetic Skyrmion Materials,
Chem.\ Rev.\ \textbf{121}, 2857 (2021).

\bibitem{Hei2011}
S.~Heinze, K.~von~Bergmann, M.~Menzel, J.~Brede, A.~Kubetzka, R.~Wiesendanger, G.~Bihlmayer, and S.~Bl\"ugel,
Spontaneous Atomic-Scale Magnetic Skyrmion Lattice in Two Dimensions,
Nat.\ Phys.\ \textbf{7}, 713 (2011).

\bibitem{Wu2021}
H.~Wu, X.~Hu, K.~Jing, and X.~R.~Wang,
Size and Profile of Skyrmions in Skyrmion Crystals,
Commun.\ Phys.\ \textbf{4}, 210 (2021).

\bibitem{finazzi2013laser}
M.~Finazzi, M.~Savoini, A.~R.~Khorsand, A.~Tsukamoto, A.~Itoh, L.~Du\`o, A.~Kirilyuk, Th.~Rasing, and M.~Ezawa,
Laser-Induced Magnetic Nanostructures with Tunable Topological Properties,
Phys.\ Rev.\ Lett.\ \textbf{110}, 177205 (2013).

\bibitem{je2018skyrmionbubble}
S.-G.~Je, P.~Vallobra, T.~Srivastava, J.-C.~Rojas-S\'anchez, T.~H.~Pham, M.~Hehn, G.~Malinowski, C.~Baraduc, S.~Auffret, G.~Gaudin, S.~Mangin, H.~B\'ea, and O.~Boulle,
Creation of magnetic skyrmion bubble lattices by ultrafast laser in ultrathin films,
Nano\ Lett.\ \textbf{18}, 7362 (2018).

\bibitem{buttner2021fluctuation}
F.~B\"uttner, B.~Pfau, M.~B\"ottcher, M.~Schneider, G.~Mercurio, C.~M.~G\"unther, P.~Hessing, C.~Klose, A.~Wittmann, K.~Gerlinger, L.-M.~Kern, C.~Str\"uber, C.~von~Korff~Schmising, J.~Fuchs, D.~Engel, A.~Churikova, S.~Huang, D.~Suzuki, I.~Lemesh, M.~Huang, L.~Caretta, D.~Weder, J.~H.~Gaida, M.~M\"oller, T.~R.~Harvey, S.~Zayko, K.~Bagschik, R.~Carley, L.~Mercadier, J.~Schlappa, A.~Yaroslavtsev, L.~Le~Guyarder, N.~Gerasimova, A.~Scherz, C.~Deiter, R.~Gort, D.~Hickin, J.~Zhu, M.~Turcato, D.~Lomidze, F.~Erdinger, A.~Castoldi, S.~Maffessanti, M.~Porro, A.~Samartsev, J.~Sinova, C.~Ropers, J.~H.~Mentink, B.~Dup\'e, G.~S.~D.~Beach, and S.~Eisebitt,
Observation of fluctuation-mediated picosecond nucleation of a topological phase,
Nat.\ Mater.\ \textbf{20}, 30 (2021).

\bibitem{zhu2024ultrafast}
K.~Zhu, L.~Bi, Y.~Zhang, D.~Zheng, D.~Yang, J.~Li, H.~Tian, J.~Cai, H.~Yang, Y.~Zhang, and J.~Li,
Ultrafast switching to zero field topological spin textures in ferrimagnetic TbFeCo films,
Nanoscale \textbf{16}, 3133 (2024).

\bibitem{Hub1998}
A.~Hubert and R.~Sch\"afer,
\textit{Magnetic Domains: The Analysis of Magnetic Microstructures}
(Springer, Berlin, Heidelberg, 1998).

\bibitem{Gil2004}
T.~L.~Gilbert,
A Phenomenological Theory of Damping in Ferromagnetic Materials,
IEEE Trans.\ Magn.\ \textbf{40}, 3443 (2004).

\bibitem{Chu2015}
A.~V.~Chumak, V.~I.~Vasyuchka, A.~A.~Serga, and B.~Hillebrands,
Magnon Spintronics,
Nat.\ Phys.\ \textbf{11}, 453 (2015).

\bibitem[]{Berg1981}
B. Berg, M. L\"uscher, 
Definition and statistical distributions of a topological number in the lattice O(3) $\sigma$-model, 
Nuclear Physics B, 190(2), 412-424 (1981).

\bibitem{Guo2024Terahertz}
X.~Guo, K.~Bertling, B.~C.~Donose, M.~Br\"unig, A.~Cernescu, A.~A.~Govyadinov, and A.~D.~Raki\'c,
Terahertz Nanoscopy: Advances, Challenges, and the Road Ahead,
Appl.\ Phys.\ Rev.\ \textbf{11}, 021306 (2024).

\bibitem{tuniz2023subwavelength}
A.~Tuniz and B.~T.~Kuhlmey,
Subwavelength Terahertz Imaging via Virtual Superlensing in the Radiating Near Field,
Nat.\ Commun.\ \textbf{14}, 6393 (2023).

\bibitem[]{qp-mm-github}
The appended animations are openly available in the Github repository at \url{ https://github.com/yuefeiliuphys/qp-mm}.

\bibitem[]{SI}
Supplemental Material summarizes the micromagnetic outcomes of the quantum printing protocol for LG modes with polarization index $s\in\{-1,0,1\}$, azimuthal index $l\in\{0,1\}$, and radial index $p\in\{0,1\}$, including the uniform initial magnetizations $\mathbf{m}_0=+\hat{\mathbf{z}}$, $\mathbf{m}_0=+\hat{\mathbf{x}}$, and $\mathbf{m}_0=-\hat{\mathbf{z}}$.


\bibitem{fu2018optical}
X.~Fu, S.~D.~Pollard, B.~Chen, B.-K.~Yoo, H.~Yang, and Y.~Zhu,
Optical Quenching of Magnetic Vortex Visualized \emph{In Situ} by Lorentz Electron Microscopy,
Microsc.\ Microanal.\ \textbf{24}(S1), 912 (2018).

\bibitem{Fan2023KZ}
Z.~Fan, A.~del~Campo, and G.-W.~Chern,
Kibble--Zurek Mechanism for Nonequilibrium Generation of Magnetic Monopoles in Spin Ices,
arXiv:2307.05267 (2023).

\bibitem{Niu2026Optica}
L.~Niu, X.~Feng, X.~Zhang, W.~Yu, Q.~Wang, Y.~Lang, Q.~Xu, X.~Chen, J.~Ma, H.~Qiu, Y.~Shen, W.~Zhang, and J.~Han,
Electric-Magnetic-Switchable Free-Space Skyrmions in Toroidal Light Pulses via a Nonlinear Metasurface,
Optica \textbf{13}, 203 (2026).



\end{thebibliography}
\end{document}